\documentclass[eqsecnum,twocolumn,pre,showpacs]{revtex4}

\usepackage{dcolumn}
\usepackage{amsfonts}
\usepackage{amsmath}
\usepackage{amssymb}
\usepackage{bm}
\usepackage{epsfig}

\begin{document}

\newcommand{\brm}[1]{\bm{{\rm #1}}}
\newcommand{\tens}[1]{\underline{\underline{#1}}}
\newcommand{\mm}{\overset{\leftrightarrow}{m}}
\newcommand{\xv}{\bm{{\rm x}}}
\newcommand{\Rv}{\bm{{\rm R}}}
\newcommand{\uv}{\bm{{\rm u}}}
\newcommand{\nv}{\bm{{\rm n}}}
\newcommand{\Nv}{\bm{{\rm N}}}
\newcommand{\ev}{\bm{{\rm e}}}
\newcommand{\av}{{\bm a}}
\newcommand{\cv}{{\bm c}}
\newcommand{\tv}{\bm{{\rm t}}}
\newcommand{\Tv}{\bm{{\rm T}}}

\title{Unconventional elasticity in smectic-$A$ elastomers}

\author{Olaf Stenull\footnote{Present address: Department of Physics and Astronomy, University of
Pennsylvania, Philadelphia, PA 19104, USA}}
\affiliation{Fachbereich Physik, Universit\"at Duisburg-Essen,
Campus Duisburg, 47048 Duisburg, Germany}

\author{T. C. Lubensky}
\affiliation{Department of Physics and Astronomy, University of
Pennsylvania, Philadelphia, PA 19104, USA}

\vspace{10mm}
\date{\today}

\begin{abstract}
\noindent We study two aspects of the elasticity of smectic-$A$
elastomers that make these materials genuinely and qualitatively
different from conventional uniaxial rubbers. Under strain applied
parallel to the layer normal, monodomain smectic-$A$ elastomers
exhibit a drastic change in Young's modulus above a threshold
strain value of about $3\%$, as has been measured in 
experiments by Nishikawa and Finkelmann [Macromol.\ Chem.\ Phys.\
{\bf 200}, 312 (1999)]. Our theory predicts that such strains induce a
transition to a smectic-$C$-like state and that it is this
transition that causes the change in elastic modulus. We calculate
the stress-strain behavior as well as the tilt of the smectic layers and the molecular orientation for
strain along the layer normal, and we compare our findings with the
experimental data. We also study the electroclinic effect in
chiral smectic-$A^\ast$ elastomers. According to experiments by Lehmann {\em et al}.\ [Nature {\bf 410}, 447 (2001)] and K\"{o}hler {\em et al}.\ [Applied Physics A {\bf 80}, 381 (2003)], this effect leads in smectic-$A^\ast$ elastomers to a giant or, respectively, at least very large lateral electrostriction. Incorporating polarization into our theory, we calculate the height change of
smectic-$A^\ast$ elastomer films in response to a lateral external
electric field, and we compare this result to the experimental
findings.
\end{abstract}


\maketitle

\section{Introduction}
\label{introduction}
\begin{figure}
\centerline{\includegraphics[width=8.4cm]{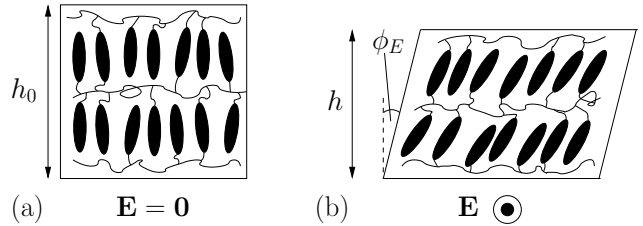}}
\caption{Electroclinic effect in Sm$A^\ast$ elastomers. (a) Without
electric field, the elastomeric film has a thickness $h_0$. (b) By
application of a lateral electric field, a tilt angle $\phi_E$ is
induced in a plane perpendicular to the field and the film thickness
decreases to a value $h$.}
\label{fig:electrostriction}
\end{figure}

Smectic elastomers~\cite{WarnerTer2003} are rubbery materials that
have the macroscopic symmetry properties of smectic liquid
crystals~\cite{deGennesProst93_Chandrasekhar92}. They possess a
plane-like, lamellar modulation of density in one direction. In
the smectic-$A$ (Sm$A$) phase, the Frank director $\brm{n}$
describing the average orientation of constituent mesogens is
parallel to the normal $\brm{N}$ of the smectic layers whereas in
the smectic-$C$ (Sm$C$) phase, it has a component in the plane of
the layers. Monodomain Sm$A$ elastomers are macroscopically
uniaxial elastomers, albeit with unusual  mechanical and electrical properties. For example, when strained along an axis normal to smectic layers,
they exhibit a reorientation of smectic layers and an associated
decrease in Young's modulus $Y_{||}$ along this axis above a
critical strain. This phenomenon, discovered experimentally by Nishikawa and Finkelmann (NF) \cite{nishikawa_finkelmann_99}, is the analog in smectic elastomers of the Helfrich-Hurault effect in
uncrosslinked smectics \cite{Helfrich-Hurault,buckling}.  Moreover, like
thermotropic chiral smectics, chiral Sm$A$
elastomers exhibit electrostriction, as depicted in
Fig.~\ref{fig:electrostriction}, in which smectic layer spacing
decreases in response to an electric field in the plane of the
layers. Early experiments~\cite{lehmann&Co_01} produced a reduction
in layer spacing by as much as 4\% at field of 1.5 MV m$^{-1}$. More
recent experiments~\cite{KohlerZen2005} produce a reduction of 1\% in
fields as high as 3 MV m$^{-1}$ in films with one free edge to
reduce mechanical stress.  Even a 1\% reduction in height is larger
than that produced in most traditional actuators. 

In this paper, we analyze the experiments just discussed using a
phenomenological model of smectic elastomers we have recently
developed
\cite{stenull_lubensky_letter_2005,stenull_lubensky_biax_smC_2006}.
This model, which generalizes the standard Lagrangian approach to
elasticity theory \cite{Landau-elas,tomsBook}, brings together in a
single rubric physics associated with the crosslinked network that
define the elastomer, the layering of the smectic phase, and the
interaction between the nematic director and both the network and
the smectic layers.  It includes energies that favor constant
smectic layer spacing and that lead to the development of shear
strain in response to stretch perpendicular to layers above a
critical stress. It also includes symmetry-permitted interactions
between $c$-ordering (molecular tilt relative to layer normal) and
shear strain that cause shear to produce $c$-ordering and vice
versa. Thus a strain along the smectic layer normal produces the 
$c$-ordering of a Sm$C$ elastomer, as well as shear strain.  We will analyze this
instability of a Sm$A$ elastomer under strain parallel to the layer
normal in terms of an induced transition to a Sm$C$ elastomer.
Comparing our numerical estimates for the stress-strain behavior and the tilt of the smectic layers to the available experimental curves, we find convincing agreement. To analyze the electrostriction experiments, we add to our model the standard
chiral couplings of the electric field to the Frank director and
layer normal.  The result is that an electric field in the plane of
the layers produces both Sm$C$ ordering and shear strain
linear in the electric field and a reduction in layer spacing
quadratic in the field.  Our numerical estimate of the
magnitude of the reduction in layer spacing is in accord with the experimental values. To our knowledge, experiments have not
sought to detect the shear strain induced by the external electric
field.

Warner and Adams (AW) \cite{adams_warner_2005} have recently
developed a theory of the elastic properties of Sm$A$ elastomers
that combines the neoclassical model of nematic elastomers
\cite{WarnerTer2003} with the compressional energetics of smectic
layers.  This theory assumes that the nematic director is locked to
the layer normal and that the elastomer is incompressible.  Our theory
is phenomenological; it is based on an expansion, consistent with
all symmetries, of the free energy in powers of the
Cauchy-Saint-Venant strain tensor \cite{Landau-elas} and the nematic
director.  The AW theory is semi-microscopic, and it applies, strictly speaking, solely to crosslinked networks of liquid crystal polymers. However, it provides a more
realistic description of these particular elastomers than does our phenomenolgical model if strains are large.  We compare the AW theory and ours in some detail, and we explain how the former theory can be extended to allow for relative tilt between the director and the layer normal. Though the details of the 
AW approach and ours us differ considerably, except for the development of
Sm$C$ order, they not surprisingly predict qualitatively
identical results.  A theory for chiral smectic elastomers was first
set up by Terentjev and
Warner~\cite{terentjev_warner_SmA_1994,terentjev_warner_SmC_1994}
using group theory. Our theory with polarization generalizes the
theory by Terentjev and Warner in a formalism that ensures
invariance with respect to arbitrary rather than infinitesimal
rotations of both the director and mass points.

The outline of our paper is as follows: Section~\ref{modelling}
briefly reviews our model for smectic
elastomers. Section~\ref{SmAelasticity} treats the behavior of Sm$A$
elastomers under strain imposed along the layer normal. Exploiting
the model reviewed in Sec.~\ref{modelling}, it derives predictions
for the stress-strain behavior as well as the layer and the molecular tilt in response to
strain. Section~\ref{electroclinicEffect} incorporates polarization
into our model of Sec.~\ref{modelling} and considers the
electroclinic effect and electrostriction.
Section~\ref{concludingRemarks} presents concluding remarks. There is
one appendix that compares our theory in some detail to the AW theory as it stands. Moreover, the appendix presents a generalization of the AW model that allows for Sm$C$ ordering.

\section{\label{modelling}Modelling Smectic Elastomers}

Let us first review briefly our model for smectic elastomers.
Smectic elastomers are, like any elastomer, permanently
crosslinked amorphous solids whose static elasticity is most
easily described in Lagrangian coordinates in which $\brm{x}$
labels a mass point in the undeformed (reference) material and
$\brm{R}(\brm{x})= \brm{x} + \brm{u}(\brm{x})$, where
$\brm{u}(\brm{x})$ is the displacement variable, labels the
position of the mass point $\brm{x}$ in the deformed (target)
material. Lagrangian elastic energies are formulated in terms of
the strain tensor $\tens{u}= \frac{1}{2} (\tens{g} -\tens{\delta})$ where $\tens{\delta}$ is the unit matrix, $\tens{g} = \tens{\Lambda}^T  \tens{\Lambda}$ is the metric tensor, and $\Lambda_{ij} =
\partial R_i/\partial x_j$ are the components of the Cauchy deformation tensor
$\tens{\Lambda}$ with $i,j,k = x,y,z$. The components of
$\tens{u}$ are thus $u_{ij} ( \xv ) = \textstyle{\frac{1}{2}}\,
(\Lambda^T_{ik}\Lambda_{kj} - \delta_{ij}) =
\textstyle{\frac{1}{2}} \, (\partial_i u_j +\partial_j u_i +
\partial_i u_k\partial_j u_k)$. Here and in the following the summation convention on repeated indices is understood. 

As in nematic elastomers, the Frank director $\nv$ interacts with
the elastic strain.  The strain $\tens{u}$ is a reference-space tensor
that transforms under the same reference-space operations that
transform $\xv$.  The director $\nv$, on the other hand, is a
target-space vector that transforms under the same target-space
operations that transform $\Rv$.  To create scalar invariants
coupling $\tens{u}$ to $\nv$, we need to convert $\nv$ to a
reference-state vector ${\tilde \nv}$.  This is accomplished with
the aid of the polar decomposition \cite{HornJoh1991} of the
deformation tensor: $\tens{\Lambda} = \tens{O} \, \tens{g}^{1/2}$,
where $\tens{g}^{1/2}$ is a symmetric reference-space tensor and
$\tens{O} = \tens{\Lambda}\,  \tens{g}^{-1/2}$ is a rotation matrix.
The components of $\tens{O}$ are $O_{ij} = \Lambda_{ik}
[g^{-1/2}]_{kj}$ where we use the notation that $[M^{\alpha}]_{ij}$
is the $ij$-component of the matrix $\tens{M}^{\alpha}$ for any
square matrix $\tens{M}$ and exponent $\alpha$ (it is understood
that $M_{ij} \equiv [M]_{ij}$).

The left and right indices in $O_{ij}$ transform, respectively,
under target- and reference-space operations, and $\tens{O}$
converts reference-space vectors to target-space vectors.  Thus,
we have $\nv = \tens{O} \, {\tilde \nv}$ and ${\tilde \nv} =
\tens{O}^T \nv$. We will express the vectors $\nv$ and ${\tilde
\nv}$ in terms of their components $c_a$ and ${\tilde c}_a$
perpendicular to the initial anisotropy axis along $z$: $n_i =
{c_a,n_z}$ where $n_z = (1- c_a^2 )^{1/2} \approx 1 - \frac{1}{2}
c_{a}^2$ and similarly for ${\tilde n}_i$, where we introduced the
notation that letters $a,b,c ...$ at the beginning of the alphabet
run over the directions $x,y$ perpendicular to $z$.  Reference-space vectors
such as $\tilde \cv$ and the vector ${\tilde \ev} = (0,0,1)$ along
the reference-space uniaxial direction \cite{n0} can be contracted
with the strain $u_{ij}$ to produce scalars like ${\tilde c}_a
u_{ab} {\tilde c_b}$ and ${\tilde c}_a u_{az}$.

We can now construct free-energy densities, which we will simply
call energies, for the various contributions to the total energy $f$
of a smectic elastomer. There is a contribution $f_{\text{net}}$ to $f$
 that is identical to the soft elastic energy of nematic elastomers including both the
familiar terms quadratic in $u_{ij}$ and ${\tilde c}_a$. The
characteristic energies scales in this free energy are set by the
volume bulk modulus $B \sim 3 \times 10^{9}$ Pa, the network shear
modulus $\mu \sim 10^6$, and moduli that scale as $\mu$ times a
factor that vanishes when the ratio $p$ of the polymer step lengths
parallel and perpendicular to the local director becomes unity.

Smectic elastomers have a layer structure with a preferred layer
spacing whose magnitude depends on the angle $\Theta$ between $\nv$
and the layer normal $\Nv$ (or, equivalently between ${\tilde \nv}$
and ${\tilde \Nv}$). Smectic order is described by a complex
mass-density-wave amplitude whose phase is $\phi = q_0[R_z - U(\Rv)]
= q_0 [z + u_z ( \xv ) - U(\Rv(\xv))]$, where $U(\Rv)$ is the
displacement field of the smectic layers, and $q_0 = 2 \pi /d$, where
$d$ is the preferred layer spacing when $\nv$ is parallel to $\Nv$.
We consider only elastomers crosslinked in the smectic phase in
which case, $U(\Rv(\xv)) = u_z ( \xv)$, and $\phi = q_0 z$.  The
energy associated with changes in the smectic layer spacing is
\begin{align}
f_{\text{layer}} = \frac{1}{2} B_{\text{sm}} q_0^{-4} [ (\nv \cdot
{\mathbf \nabla} \phi )^2 - q_0^2 ]^2 , 
\end{align}
where $B_{\text{sm}}$ is the
smectic compression modulus with a value of order $10^7$ Pa deep in
the smectic phase though it vanishes as the nematic phase is
approached. $f_{\text{layer}}$ favors a layer spacing of $d' =
d/\cos \Theta$, where as before $\Theta$ is the angle between $\nv$
and $\Nv$.

Finally, there are interactions favoring ${\tilde \nv}$ parallel to
${\tilde \Nv}$ in the smectic-$A$ phase and tilted relative to
${\tilde \Nv}$ in the smectic-$C$ phase: 
\begin{align}
f_{\text{tilt}} = \frac{1}{2} r_t \sin^2 \Theta + \frac{1}{4} v_t \sin^4 \Theta \, . 
\end{align}
The energy $f_{\text{layer}} + f_{\text{tilt}}$ is the generalization to smectic elastomers of the
Chen-Lubensky model \cite{Chen-Lubensky} for Sm$A$-Sm$C$-nematic
phase behavior in uncrosslinked liquid crystals. Its  coefficient $r_t$ is linear in the deviation of the temperature $T$ from the transition temperature $T_c$ between the Sm$A$ and Sm$C$ phases, $r_t = \alpha |T - T_c|$. The coefficient $v_t$ is essentially independent of temperature. Experiments by Brehmer, Zentel, Gie{\ss}elmann, Germer and Zungenmaier (BZGGZ)~\cite{brehmer&Co_2006} on smectic elastomers measured the values of the coefficients of a tilt energy similar to our $f_{\text{tilt}}$ (however, their tilt energy is somewhat more general in that it also accounts for polarization effects which we will discuss later). A short calculation translates their experimental values into estimates for our $\alpha$ and $v_t$: $\alpha \sim 1.3 \times 10^4$ Pa/K and $v_t \sim 5 \times 10^5$ Pa. At a temperature of 20K above the transition temperature, $r_t$ is or order $10^5$ Pa. More recently, similar experiments were performed on liquid smectics by Archer and Dierking (AD)~\cite{ArcherDie2005}. Their results lead to $\alpha \sim 4 \times 10^4$ Pa/K and $v_t \sim 10^6$ Pa implying that $r_t \sim 10^6$ Pa at 20K above the transition temperature. Though measured for liquid smectics, the AD values should be of some relevance to elastomeric smectics, and we will use them in the following in addition to the BZGGZ values in order to put the estimates that we are going to make on a broader basis. 

The energy of a smectic elastomer $f= f_{\text{net}} +
f_{\text{layer}}+ f_{\text{tilt}}$ can now be expressed as a sum of
a harmonic uniaxial elastic contribution $f_{\text{uni}}$ depending only on
$u_{ij}$, a contribution $f_{\text{nonlin}}$ depending only on
$u_{ij}$ that collects the relevant nonlinear terms,
a contribution $f_c$ describing the energy associated with
the formation of non vanishing $c$-order described by ${\tilde
c}_a$, and a contribution $f_{\text{coupl}}$ arising from the
coupling of $u_{ij}$ and ${\tilde c}_a$:
\begin{equation}
\label{completeEn}
f= f_{\text{uni}} + f_{\text{nonlin}} + f_c + f_{\text{coupl}} .
\end{equation}
The uniaxial elastic energy to harmonic order in strains
is
\begin{align}
\label{uniEn}
f_{\text{uni}} &= \textstyle{\frac{1}{2}} \, C_1\, u_{zz}^2 + C_2
\, u_{zz} u_{ii} +  \textstyle{\frac{1}{2}} \, C_3\, u_{ii}^2
\nonumber \\
&+ C_4 \, \hat{u}_{ab}^2 + C_5 \,u_{az}^2 ,
\end{align}
where
\begin{equation}
\hat{u}_{ab}=u_{ab}-\ \textstyle{\frac{1}{2}}\, \delta_{ab} u_{cc}
\label{eq:hatu}
\end{equation}
is the two-dimensional symmetric, traceless strain tensor. The nonlinear energy reads 
\begin{align}
\label{nonlinearEn}
 f_{\text{nonlin}} =  - B_1 u_{zz} u_{az}^2 +
B_2 (u_{az}^2)^2 .
\end{align}
The $c$-director energy is
\begin{equation}
f_c = \textstyle{\frac{1}{2}} \, r \, \tilde{c}_a^2
+\textstyle{\frac{1}{4}} \, v \, (\tilde{c}_a^2)^2 ,
\end{equation}
and the coupling energy is
\begin{align}
\label{fCoupl}
f_{\text{coupl}} &=  \lambda_1\,  \tilde{c}_a^2 u_{zz} + \lambda_2
\, \tilde{c}_a^2 u_{ii}  +  \lambda_3 \, \tilde{c}_a \hat{u}_{ab}
\tilde{c}_b
\nonumber \\
&+  \lambda_4 \, \tilde{c}_a u_{az} +  \lambda_5 \, u_{zz} \tilde{c}_a u_{az}  \, .
\end{align}
Table~\ref{table1} reviews the relations between the original elastic constants of $f_{\text{net}}$,
$f_{\text{layer}}$ and $f_{\text{tilt}}$ and the effective elastic constants featured in Eq.~(\ref{uniEn}) and Eqs.~(\ref{nonlinearEn}) to (\ref{fCoupl}). As mentioned earlier, the parameter $p$ appearing in Table~\ref{table1} is the anisotropy ratio. In the work of NF, the samples showed at the transition temperature a spontaneous stretch along the director of about $12\%$. When making estimates, we will view $p\approx 1.1$ as a typical value. Based on the relations given in Table~\ref{table1}, we deduce that the hierarchy of magnitudes of the effective elastic constants is:  $C_3 \sim10^9$ Pa,  $C_1 \sim B_1 \sim B_2 \sim v \sim \lambda_1 \sim \lambda_5 \sim 10^7$ Pa, $C_2 \sim C_4 \sim C_5 \sim r \sim \mu \sim 10^6$ Pa, $\lambda_3 \sim - 10^5$Pa and  $\lambda_2 \sim -10^4$ Pa. Our estimates for $r$ and $\lambda_4$ depend noticeably on whether we use the BZGGZ or the AD values: we expect $r \sim 10^5$ Pa  and $\lambda_4 \sim -10^5$ Pa based on BZGGZ and $r \sim \lambda_4 \sim 10^6$ Pa based on AD. It is worth commenting on the sign of $\lambda_4$. Our estimates of its value have different signs for the two physical systems for which we have data. This is possible because the two contributions to $\lambda_4$ have opposite signs. The second contribution, $r_t$, which is positive, arises because $\Theta \approx \tilde{c}_a + u_{az}$, and the $\frac{1}{2} r_t \sin^2 \Theta$-term favors $\tilde{c}_a = - u_{az}$. The first term, $-\mu (p^2 -1)/p$, describes the rotation of $\tilde{c}_a$ in response to the imposition of  a shear $u_{az}$. It is negative and it favors a $\tilde{c}_a$ with the same sign as  $u_{az}$.
\begin{table*}
\caption{\label{table1} Contributions to coefficients in $f$ from
$f_{\text{net}}$, $f_{\text{layer}}$ and $f_{\text{tilt}}$ with $a= \mu\,(p-1)^2/(2p)$.}
\begin{tabular}{|l|c|c|c|c|c|c|c|c|c|c|c|c|c|c|}
\hline
     &$C_1$ & $C_2$ & $C_3$ & $C_4$ & $C_5$ & $B_1$ & $B_2$
     &$\lambda_1 $& $\lambda_2$ & $\lambda_3$ & $\lambda_4$ &
     $\lambda_5$ & $r$ & $v$ \\
     \hline
  net & $3 \mu$ & $-\mu$ & $4B-\mu$ & $\mu$ & $\frac{1}{2} \mu
  \frac{(p+1)^2}{p}$ & $ a$ & $-a$ & $\mu \frac{2p^2-p-1}{2p}$ & $-\mu \frac{p -
  1}{2p}$ & $-\mu \frac{p - 1}{p}$ & $-\mu\frac{p^2-1}{p}$ & $-\frac{3}{2} a$ & $ 2a$ & 0
  \\ \hline
  layer& $4B_{\text{sm}}$& 0 & 0 & 0 & 0 & $
  6 B_{\text{sm}}$ & $\frac{9}{2}B_{\text{sm}}$ & $2 B_{\text{sm}}$ & 0 & 0 & 0
  & $ 4 B_{\text{sm}}$ & 0 & $2 B_{\text{sm}}$
  \\ \hline
  tilt& 0 & 0 & 0 & 0 & $\frac{1}{2} r_t$ & $\frac{1}{2} r_t$ &
  $\frac{1}{2} r_t + \frac{1}{4} v_t$ & 0 & 0 & 0 & $r_t$ & $
  -\frac{1}{2} r_t$ & $ r_t $ & $v_t$
  \\ \hline
\end{tabular}
\end{table*}

Several observations about $f$ are in order.  First, ${\tilde c}_a$
and $u_{az}$ are coupled at linear order via the $\lambda_4$ term.
Thus, if a nonzero $u_{az}$ develops, it will be accompanied by the
development of a nonzero ${\tilde c}_a$ and vice-versa. Negative $\lambda_4$ favors mechanical tilt and mesogenic tilt in the same direction whereas positive $\lambda_4$ favors opposite tilt directions. If, for example, $u_{zz} =0$ and $u_{az} \neq 0$, then $\tilde{c}_a$ relaxes to $\tilde{c}_a = - (\lambda_4/r) u_{az}$ to minimize $f$. This leads to and effective or renormalized renormalized value of the shaer modulus $C_5$,
\begin{align}
\label{C5ren}
C_5^R = C_5 - \frac{\lambda_4^2}{2r} 
\end{align}
If, to give another example, $u_{zz} =0$ and $\tilde{c}_a \neq 0$, $u_{az}$ relaxes to $u_{az} = - \lambda_4/(2 C_5) \tilde{c}_a$ and $r$ is renormalized to $r_R = r  - \lambda_4^2/(2C_5)$. Note that the value of the renormalized elastic constants $C_5^R$ and $r_R$ does not depend on the sign of $\lambda_4$.
Second, an externally imposed strain $u_{zz}$ reduces the coefficient of
$u_{az}^2$ via the $B_1$ term.  This leads to the elastomer version
of the Helfrich-Hurault \cite{Helfrich-Hurault,buckling}
instability, which we will discuss in Sec.~\ref{SmAelasticity}.
Finally, if ${\tilde c}_a$ becomes nonzero, there will be an
accompanying decrease in $u_{zz}$ and the smectic layer spacing
proportional to ${\tilde c}_a^2$ via the ${\tilde c}_a^2 u_{zz}$-term
with positive coupling constant $\lambda_1$.  The reduction of layer
spacing in response to an external electric field that tilts the
director in chiral systems will be the subject of
Sec.~\ref{electroclinicEffect}

\section{Elasticity of Smectic-$A$ elastomers under strain imposed along the layer normal}
\label{SmAelasticity}
\begin{figure}
\centerline{\includegraphics[width=8.8cm]{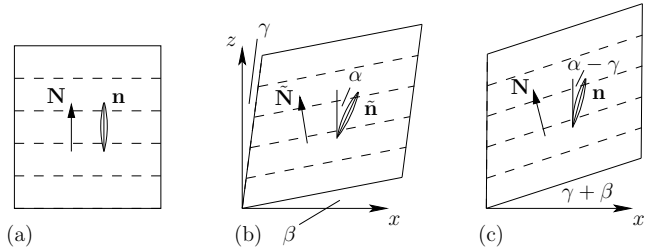}}
\caption{Schematic representation of distortions in the $xz$-plane
induced by the Sm$A$-to-Sm$C$ transition: (a) undistorted Sm$A$
phase, (b) Sm$C$ phase with a symmetric deformation tensor
$\tens{\Lambda}_S$, and (c) Sm$C$ phase with $\Lambda_{xz} = 0$ and
$\Lambda_{zx} > 0$.}
\label{fig:strains}
\end{figure}

Before we start with our analysis, it is useful to clarify the
experimental conditions. In the experiments of NF~\cite{nishikawa_finkelmann_99}, the Sm$A$ sample is
subjected to an external uniaxial stress normal to the smectic
layers.  This stress stretches the sample, producing a deformation
tensor $\tens{\Lambda}$ with $\Lambda_{zz}>1$. If there is a
transition to the Sm$C$ phase, the sample will undergo a shape
change, as sketched in Fig.~\ref{fig:strains}, from a rectangular
parallelepiped to a non-rectangular one with two parallel
non-rectangular faces.  The external stress is applied along the
space-fixed $z$-axis and causes the target space vector $\Rv$ to
stretch along the $z$-axis. This defines a preferred orientation in the
target space and therefore breaks rotational invariance in this space. In an
idealized experiment, the external stress will stretch the sample
parallel to the $z$-axis in the target space only, regardless of the
shape of the non-rectangular faces.  Thus, in the Sm$C$ phase, the
configuration of the sample will be that (shown in Fig.\
\ref{fig:strains}(c)) of a simple shear in which $\Lambda_{zx}>0$
and $\Lambda_{xz} = 0$. In a real experiment, boundary conditions
prevent a distortion with a spatially uniform value of
$\Lambda_{zx}$, and the system breaks up into oppositely tilted
domains.  We will ignore this effect in our analysis. Thus, the
general form for the deformation tensor under uniaxial stress along
the target-space $z$ axis is
\begin{equation}
\label{formOfLambda}
\tens{\Lambda}= \left(
\begin{array}{ccc}
 \Lambda_{xx} & 0 & 0 \\
 0 & \Lambda_{yy} & 0 \\
 \Lambda_{zx} & 0 & \Lambda_{zz}
\end{array}
\right) ,
\end{equation}
where $\Lambda_{zx} = 0$ in the Sm$A$ phase.  This then implies a
strain tensor
\begin{equation}
\label{experimentalU}
\tens{u} = \frac{1}{2} \left(
\begin{array}{ccc}
 \Lambda_{xx}^2 + \Lambda_{zx}^2 -1 & 0 & \Lambda_{zx} \Lambda_{zz}
 \\
 0 & \Lambda_{yy}^2 - 1 & 0 \\
 \Lambda_{zx} \Lambda_{zz} & 0 & \Lambda_{zz}^2 - 1
\end{array}
\right) .
\end{equation}
Thus, the $zz$-component of the strain tensor, $u_{zz} =
(\Lambda_{zz}^2 -1)/2$, depends only on the imposed strain
deformation $\Lambda_{zz}$ in both phases. The formation of the
Sm$C$ phase is signalled by the development of a nonzero shear
strain $u_{xz} = \Lambda_{zx} \Lambda_{zz}/2$.

As we indicated earlier, the imposition of a sufficiently large
positive strain $u_{zz}$ induces a transition to a new phase with a
nonvanishing shear strain $u_{az}$. Because of the linear coupling
$\lambda_4 \tilde{c}_a u_{az}$ between strain and $c$-director in
$f_{\rm coupl}$, director tilt develops along with shear strain, and
the phase with nonzero $u_{az}$ can be viewed as a strain-induced
Sm$C$ phase. Effective energies in terms of either $u_{az}$ or
$\tilde{c}_a$ provide equally valid descriptions of the transition.
Here we will describe the transition in terms of an effective energy
expressed in terms of $\tilde{c}_a$.  To arrive at this energy, we
must relax the variables $u_{ii}$, ${\hat u}_{ab}$ and $u_{az}$ to
their equilibrium values in the presence of nonvanishing
$\tilde{c}_a$ and $u_{zz}$.  The equations for $\partial f/\partial
u_{ii} = 0$ and $\partial f /\partial \hat{u}_{ab} = 0$ are linear
and are easily solved for $u_{ii}$ and ${\hat u}_{ab}$:
\begin{subequations}
\begin{align}
u_{ii} & = - \frac{C_2}{C_3} u_{zz}- \tau \tilde{c}_a^2 ,
\label{eq:u-ii1}\\
\hat{u}_{ab} & = - \omega \left(\tilde{c}_a \tilde{c}_b -
\textstyle{\frac{1}{2}}\delta_{ab} \tilde{c}_c^2 \right) ,
\label{eq:u-ab1}
\end{align}
\end{subequations}
where
\begin{equation}
\tau = \lambda_2/C_3\qquad \text{and} \qquad \omega =
\lambda_3 /(2\, C_4) .
\end{equation}
The equation of state for $u_{az}$ is nonlinear:
\begin{equation}
\frac{\partial f}{\partial u_{az}}  = 2 C_5(u_{zz}) u_{ab} + 4 B_2
u_{bz}^2 u_{az} + \lambda_4 (u_{zz} )\tilde{c}_a   ,
\label{eq:u-az1}
\end{equation}
where
\begin{subequations}
\begin{align}
C_{5} ( u_{zz} ) & =  C_5 - B_1 u_{zz} \, ,
 \\
\lambda_4( u_{zz} ) & =  \lambda_4 + \lambda_5 u_{zz} \, .
\end{align}
\end{subequations}
Since we are interested in properties near the transition, we solve
Eq.~(\ref{eq:u-az1}) for $u_{az}$ to third-order in $\tilde{c}_a$:
\begin{equation}
u_{az} = - \rho(u_{zz}) \tilde{c}_a + 2 B_2 \, \frac{\rho^3(u_{zz})}{C_5 (u_{zz})} \, 
\tilde{c}_b^2 \tilde{c}_a ,
\label{eq:u-az2}
\end{equation}
where
\begin{equation}
\rho(u_{zz}) = \frac{\lambda_4 (u_{zz})}{2 C_{5} ( u_{zz} )} .
\end{equation}
Inserting Eqs.~(\ref{eq:u-ii1}), (\ref{eq:u-ab1}) and
(\ref{eq:u-az2}) into the free energy of Eq.~(\ref{completeEn}), we
obtain an effective energy expressed as a function of $\tilde{c}_a$
and $u_{zz}$ (the latter being fix externally),
\begin{equation}
f_{\rm eff} = \textstyle{\frac{1}{2}} C_{1}^R u_{zz}^2 +
\textstyle{\frac{1}{2}} r_R(u_{zz}) \tilde{c}_a^2 +
\textstyle{\frac{1}{4}}v_R(u_{zz}) \tilde{c}_a^2 \tilde{c}_b^2 ,
\label{eq:f-eff}
\end{equation}
where
\begin{subequations}
\begin{align}
C_{1}^R & = C_1 - C_2^2/C_3 \, ,\\
r_R(u_{zz}) & =  r + 2( \lambda_1  - C_2 \tau) u_{zz} - 2 C_{5}(u_{zz} )\rho^2 (u_{zz}) \, , \\
v_{R}(u_{zz}) & =  v- 2 C_4 \omega^2 - 2 C_3 \tau^2  + 4 B_2 \rho^4
(u_{zz}) \, .
\end{align}
\end{subequations}

We can now use $f_{\rm eff}$ to analyze the transition to the
Sm$C$ phase induced by the uniaxial strain $u_{zz}$. The
term $2 C_{5}(u_{zz} )\rho^2 (u_{zz}) = \lambda_4^2( u_{zz})/2
C_{5}( u_{zz} )$ in $r_R(u_{zz})$ diverges at $u_{zz} = C_5/B_1$ and guarantees that $r_R (u_{zz} )$
will pass through zero upon increasing $u_{zz}$. $u_{zz}^c$, the critical value of $u_{zz}$ at which
$r_R = 0$, is given by
\begin{equation}
u_{zz}^c = \Delta^{-1} \Big[  \Xi - \sqrt{ \Xi^2 - 2 r C_5^R \Delta}\Big] ,
\end{equation}
with $C_5^R$ as given in Eq.~(\ref{C5ren}), and where $\Delta = 4 B_1 (\tau C_2 - \lambda_1) - \lambda_5^2$ and $\Xi = r B_1 + 2 \tau C_2 C_5 - 2 C_5 \lambda_1 +  \lambda_4 \lambda_5$.
Both, the BZGGZ and the AD values lead to $u_{zz}^c \approx 0.03$
 in good agreement with the experimental estimate of
$0.034$ \cite{nishikawa_finkelmann_99}. In the vicinity of the
critical point determined by $u_{zz}^c$, we can expand $r_R(u_{zz})$
and $g_R(u_{zz})$ to lowest order in $\delta u_{zz} = u_{zz} -
u_{zz}^c$:
\begin{subequations}
\begin{align}
r_R(u_{zz}) & =  - b \, \delta u_{zz} \, ,
 \\
v_R(u_{zz}) & =  v_R(u_{zz}^c )\equiv v_R^c ,
\end{align}
\end{subequations}
where $b = - \partial r_R(u_{zz})/\partial u_{zz} |_{u_{zz} =
u_{zz}^c} > 0$. Choosing our coordinate system
so that the $c$-director induced by the $u_{zz}$ strain lies along
the $x$-direction we obtain $\tilde{c}_y^0 =  0$ and
\begin{align}
\tilde{c}_x^0 = \left\{
\begin{array}{ccc}
0& \quad \mbox{for} \quad & u_{zz} < u_{zz}^c \, ,
\\
\pm s \sqrt{\delta u_{zz}}& \quad \mbox{for} \quad& u_{zz} >
u_{zz}^c \ ,
\end{array}
\right.
\label{eq:c0x}
\end{align}
where $s=\sqrt{b/v_R^c}$. The director in the
Sm$C$ phase under a symmetric shear, i.e.\ in the geometry shown
in Fig.~\ref{fig:strains}(b), is therefore $\tilde{\brm{n}} =
(\sin\alpha, 0 , \cos \alpha )$, where $\sin \alpha =
\tilde{c}_x^0$. Using the BZGGZ and the AD values, we estimate $s \approx 0.75$ and $s \approx 0.8$, respectively. 

From the equilibrium values of the $c$-director,  it is a
straightforward exercise to determine the equilibrium strain
tensor $\tens{u}^0$ from Eqs.~(\ref{eq:u-ii1}), (\ref{eq:u-ab1}),
and (\ref{eq:u-az2}). Consistent with Eq.~(\ref{experimentalU}),
there is no shear induced by $u_{zz}$ in the $xy$-plane, $u_{xy}^0
=  0 $, and there is no shear induced in the $yz$-plane, $u_{yz}^0
=  0$. There is, however, a shear in the $xz$-plane, which
according to Eq.~(\ref{eq:u-az2}) and (\ref{eq:c0x}) is
\begin{align}
u_{xz}^0 = \left\{
\begin{array}{ccc}
0& \quad \mbox{for} \quad & u_{zz} < u_{zz}^c \, ,
\\
\mp \rho_c s \, \sqrt{\delta u_{zz}}& \quad \mbox{for} \quad& u_{zz} >
u_{zz}^c \ ,
\end{array}
\right.
\label{eq:u0xz}
\end{align}
to lowest order in $\delta u_{zz}$ with $\rho_c \equiv \rho(u_{zz}^c)$. The BZGGZ and AD values, respectively, lead to $\rho_c \approx 1.2$ and  $\rho_c \approx 1.1$.

\begin{figure}
\centerline{\includegraphics[width=7.5cm]{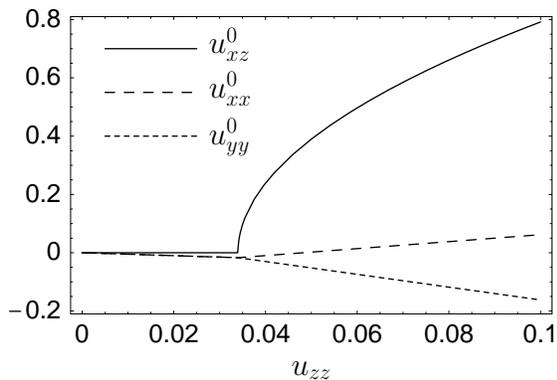}}
\caption{Dependence (schematic) of the non-zero equilibrium strain
components on $u_{zz}$.}
\label{equilibriumComponents}
\end{figure}

The equilibrium values for extensional strains follow from
Eqs.~(\ref{eq:u-ii1}), (\ref{eq:u-ab1}) and $u_{ii} = u_{aa} +
u_{zz}$.  To lowest order in $\delta u_{zz}$,
\begin{subequations}
\label{equiU11CompleteModel}
\begin{align}
u_{xx}^0 &=  -\frac{1}{2} \bigg[  1 + \frac{C_2}{C_3} \bigg] u_{zz}
\quad \mbox{for} \quad  u_{zz} < u_{zz}^c \, ,
\\
u_{xx}^0 &= -\frac{1}{2} \bigg[  1 + \frac{C_2}{C_3} \bigg] u_{zz}^c
- \frac{1}{2} \bigg[  1 + \frac{C_2}{C_3}
\nonumber \\
&+ (\tau + \omega) a^2 \bigg] \delta u_{zz} \quad
\mbox{for} \quad  u_{zz} < u_{zz}^c\, ,
\end{align}
\end{subequations}
and $u_{yy}^0$ is given by Eq.~(\ref{equiU11CompleteModel}) with
$\omega$ replaced by $-\omega$. The parameter
$\omega = \lambda_3/(2 C_5) = - (\mu (p-1)/p) [r_t + \mu
(p-1)^2/p]$ is negative and of order $-0.05$.
Figure~\ref{equilibriumComponents} depicts the equilibrium strain
components on $u_{zz}$ schematically.

The quantities ${\tilde c}_x^0$ and $u_{xz}^0$ determine the
equilibirum angle $\Theta^0$ between the layer normal and the
director. To lowest order,
\begin{align}
\Theta^0 &= {\tilde c}_x^0 + u_{xz}^0 = [1- \rho_c] {\tilde
c}_x^0   .
\end{align}
The coefficient $1- \rho_c$ of ${\tilde c}_x^0$ on the right hand side is negative and of the order of $-0.2$ or $-0.1$, depending on whether we use the BZGGZ or the AD values. The small but non-zero $1- \rho_c$ implies that the tilt angles of the layer normal and the director are distinct and that, however, their difference will be small for $u_{zz}$ not too far above $u_{zz}^c$. We will revisit the tilt angles of the layer normal and the director further below.

Now we turn to the effective elastic energy density as a function
of $u_{zz}$ only when the other strain components and $\tilde{\brm{c}}$ have relaxed to
their equilibrium values. Plugging $\tilde{c}_a^0$, into
Eq.~(\ref{eq:f-eff}), we obtain to quadratic order in $u_{zz}$
\begin{subequations}
\label{effectiveCompleteModel}
\begin{align}
\label{effective1CompleteModel}
f  =  \textstyle{\frac{1}{2}} C_1^R u_{zz}^2 \quad \mbox{for} \quad
& u_{zz} < u_{zz}^c  \, ,
\end{align}
and
\begin{align}
\label{effective2CompleteModel}
f  &=  \textstyle{\frac{1}{2}} C_1^R \left( u_{zz}^c + \delta u_{zz}
\right)^2 -  \textstyle{\frac{1}{2}} D_1\, (\delta u_{zz})^2 \quad
\mbox{for} \quad  u_{zz} > u_{zz}^c \, ,
\end{align}
\end{subequations}
where $D_1=(1/2)b^2/v_R^c$. Note that the two branches
(\ref{effective1CompleteModel}) and (\ref{effective2CompleteModel})
match at $u_{zz}^c$ as they should.

Next, we address the relation between stress and strain.  The
engineering or first Piola-Kirchhoff stress measured in experiments
is $\sigma^{\text{eng}}_{ij} = \Lambda_{ik} \sigma_{kj}$, where
$\sigma_{kj} = \partial f/\partial u_{kj}$ is the symmetric second
Piola-Kirchhoff stress tensor. The system reaches equilibrium at
fixed $u_{zz}$ with respect to all other components of $u_{ij}$,
i.e., $\sigma_{ij} = 0$ for $ij \neq zz$. Thus, the $zz$ component
of the engineering stress tensor is $\Lambda_{zz} \sigma_{zz}$, and
\begin{subequations}
\label{effectiveStressCompleteModel}
\begin{align}
\label{effectiveStress1CompleteModel}
\sigma_{zz}^{\text{eng}}  = C_1^R \Lambda_{zz} u_{zz} \quad
\mbox{for} \quad & u_{zz} < u_{zz}^c  \, ,
\end{align}
and
\begin{align}
\label{effectiveStress2CompleteModel}
\sigma_{zz}^{\text{eng}}  =   C_1^R u_{zz}^c \Lambda_{zz} + [C_1^R
-D_1]  \Lambda_{zz} \delta u_{zz} \quad \mbox{for} \quad u_{zz}>
u_{zz}^c \, .
\end{align}
\end{subequations}
For the deformation tensor of Eq.(\ref{formOfLambda}) and small
strain, $\Lambda_{zz} = \sqrt{1 + 2 u_{zz}} \approx 1 + u_{zz}$.
Thus, if the critical strain $u_{zz}^c$ is small as our estimates
indicate,
\begin{subequations}
\label{linearizedStress}
\begin{align}
\label{linearizedStressa}
\sigma_{zz}^{\text{eng}} \approx C_1^R u_{zz}  \quad \mbox{for}
\quad & u_{zz} < u_{zz}^c .
\end{align}
For $u_{zz} > u_{zz}^c$ and small $\delta u_{zz}$,
\begin{align}
\label{linearizedStressb}
\sigma_{zz}^{\text{eng}} \approx C_1^R u_{zz}^c + [C_1^R -D_1]
\delta u_{zz}  \quad \mbox{for} \quad & u_{zz} > u_{zz}^c ,
\end{align}
\end{subequations}
provided that $C_1^R \neq D_1$, and hence the Young's modulus $Y_\parallel$ is given by
\begin{align}
Y_\parallel = \left\{
\begin{array}{ccc}
C_1^R& \quad \mbox{for} \quad & u_{zz} < u_{zz}^c \, ,
\\
C_1^R - D_1& \quad \mbox{for} \quad& u_{zz} >
u_{zz}^c \ ,
\end{array}
\right.
\end{align}
in this case. Adams and Warner in fact find $C_1^R
= D_1$ in their model, and we obtain the same result in our
treatment if we lock $\nv$ to $\Nv$ and set $f= f_{\text{net}}+
f_{\text{layer}}$.  This limit does not follow cleanly from the
model in Eq.~(\ref{completeEn}) and the parameters of Table
{\ref{table1} because that model does not include all of the
fourth-order terms in the free energy of the original model defined
$f=f_{\text{net}} + f_{\text{layer}} + f_{\text{tilt}}$ . If $C_1^R
= D_1$, then $\sigma_{zz}^{\text{eng}} = C_1^R u_{zz}^c
\Lambda_{zz}$ so that the leading terms for $u_{zz}^c \ll \delta
u_{zz} \ll 1$ are
\begin{align}
\label{linearizedStressc}
\sigma_{zz}^{\text{eng}} \approx C_1^R u_{zz}^c + C_1^R u_{zz}^c
\delta u_{zz}  \quad \mbox{for} \quad & u_{zz} > u_{zz}^c ,
\end{align}
implying that
\begin{align}
Y_\parallel = \left\{
\begin{array}{ccc}
C_1^R& \quad \mbox{for} \quad & u_{zz} < u_{zz}^c \, ,
\\
C_1^R  u_{zz}^c & \quad \mbox{for} \quad& u_{zz} >
u_{zz}^c \ ,
\end{array}
\right.
\end{align}
if $C_1^R= D_1$. Equations~(\ref{linearizedStress}) and (\ref{linearizedStressc}) reveal that
$\sigma_{zz}^{\text{eng}}$ as a function of $u_{zz}$ will consist of
two straight lines of different slope (Young's modulus) meeting at $u_{zz} =
u_{zz}^c$. If $C_1^R = D_1$, then the ratio of the Young's modulus for $u_{zz} > u_{zz}^c$ to that for
$u_{zz} < u_{zz}^c$ is simply $u_{zz}^c$. Our full
equations~(\ref{effectiveStressCompleteModel}) show that there are
nonlinear corrections to $\sigma_{zz}^{\text{eng}}$ both above and
below $u_{zz}^c$.  Since the experimental value of $u_{zz}^c \approx
0.034$ is much less than one, the nonlinear corrections below
$u_{zz}^c$ are unimportant. Those above $u_{zz}^c$ vanish or are
unimportant if $C_1^R$ and $D_1$ are equal or nearly so.
Figure~\ref{SmAstressStrain} shows the linearized stress-strain
behavior along with experimental data. Both,
Eq.~(\ref{linearizedStressb}) and (\ref{linearizedStressc}) are used
in the plots and both produce excellent agreement with the
experimental data.
\begin{figure}
\centerline{\includegraphics[width=8.4cm]{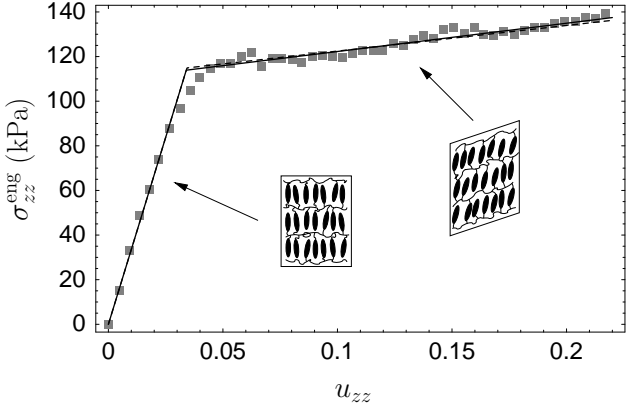}}
\caption{The engineering stress $\sigma_{zz}^\text{eng}$ as a
function of strain $u_{zz}$ of a monodomain Sm$A$ elastomer. The
squares symbolize experimental data by Nishikawa and
Finkelmann~\cite{nishikawa_finkelmann_99}. Above a threshold strain
of about $3\%$ the elastomer is in a Sm$C$-like phase with an
elastic modulus much lower than that of the original Sm$A$ phase
below the threshold. The continuous lines stem from
Eqs.~(\ref{linearizedStressa}) and (\ref{linearizedStressb}) with
$u_{zz}^c = 0.034$, $C_1^R  = 3.5 \times 10^7 \, \mbox{Pa}$, and $D_1
= 3.223 \times 10^7 \, \mbox{Pa}$. The dashed lines stem from
Eqs.~(\ref{linearizedStressa}) and (\ref{linearizedStressc}) with
$u_{zz}^c = 0.0343$ and $C_1^R  = 3.5 \times 10^7 \, \mbox{Pa}$. The
continuous and dashed lines lie almost on top of each other and are
therefore hard to distinguish.}
\label{SmAstressStrain}
\end{figure}

For completeness we note that, if we use our strain-only model for
the Sm$A$-to-Sm$C$ transition~\cite{stenull_lubensky_letter_2005},
we find, to leading order in $\delta u_{zz}$, essentially the same
equilibrium strain tensor, the same effective elastic energy, and
the same stress-strain relation as AW. Obvious differences reside in
the different definitions of the elastic constants used in both
models. These differences are of course qualitatively unimportant.

Next we calculate the tilt angle $\varphi$ of the layer normal in
response to $u_{zz}$. As illustrated in Fig.~\ref{fig:strains}, the
angle that $\brm{N}$ makes with the $z$ axis is
\begin{equation}
\varphi = \beta + \gamma \, ,
\end{equation}
where $\beta$ is the angle that $\tilde{\brm{N}}$ makes with the $z$-axis and $\gamma$
is the clockwise angle about the $y$-axis through which the sample
under symmetric shear has to be rotated to bring it into the
configuration with $\Lambda_{xz} = 0$. In terms of the components of the symmetric deformation tensor $\tens{\Lambda}_S$, these angles are given by
\begin{subequations}
\begin{align}
\beta &= \tan^{-1} \left( \frac{\Lambda_{Szx}^0}{\Lambda_{Sxx}^0}  \right) \approx \frac{u_{xz}^0}{1 - \frac{1}{2} (1 + C_2/C_3) u_{zz}^c} \, ,
\\
\gamma &= \tan^{-1} \left( \frac{\Lambda_{Szx}^0}{\Lambda_{Szz}}  \right) \approx \frac{u_{xz}^0}{1 + u_{zz}^c} \, .
\end{align}
\end{subequations}
We find that $\varphi$ vanishes as $\delta
u_{zz} \rightarrow 0$, and near $\delta u_{zz} = 0$, its expansion
to order $(\delta u_{zz})^{3/2}$ is
\begin{equation}
\label{resVarphi}
\varphi = \left\{
\begin{array}{ccc}
0& \quad \mbox{for} \quad & u_{zz} < u_{zz}^c \, ,
\\
 A_1 \, (\delta u_{zz})^{1/2} + A_3 \, (\delta u_{zz})^{3/2}
 & \quad \mbox{for} \quad& u_{zz} > u_{zz}^c \ .
\end{array}
\right.
\end{equation}
The value of $A_1$ is determined entirely by the lowest order
expressions for $u_{xz}^0$
[Eq.~(\ref{eq:u0xz})] and $u_{xx}^0$ [Eq.~(\ref{equiU11CompleteModel})],
\begin{equation}
A_1  =  s  \rho_c \,  \frac{2 + \frac{1}{2} (1 - C_2/C_3) u_{zz}^c}{[1 + u_{zz}^c][1 - \frac{1}{2} (1 + C_2/C_3) u_{zz}^c]} \, .
\label{eq:A_1}
\end{equation}
$A_3$ has contributions both from the expansions of  $\tan^{-1}$ to third order in its arguments, and from $(\delta u_{zz})^{3/2}$ contributions to $u_{xz}^0$ and $u_{xx}^0$, which we have not calculated. Both, the BZGGZ and AD values, lead to $A_1 \approx 1.8$. Next, we compare our result~(\ref{resVarphi}) and our estimate for $A_1$ with the experimental data by NF, see Fig.~(\ref{angleVsStrain}). For this comparison it is important to note that the expansion~(\ref{resVarphi}) with $A_1$ given by Eq.~(\ref{eq:A_1}) expresses $\varphi$ in its natural unit, radian. The experimental curve by NF as shown Fig.~(\ref{angleVsStrain}), on the other hand, measures $\varphi$ in degrees. Converting the above estimate for $A_1$ from radian to degrees we get the estimate $A_1 \approx 1.8 \times 360/(2 \pi) \approx 100$. Fitting our result~(\ref{resVarphi}) to the experimental curve, we obtain $u_{zz}^c = 0.034$, $A_{1} = 90$ and $A_{3} =-49$. The fitted analytical and the experimental curve agree remarkably well, and the best-fit value $A_{1} = 90$ is certainly of the same order of magnitude as our estimate $A_{1} \approx 100$
\begin{figure}
\centerline{\includegraphics[width=8cm]{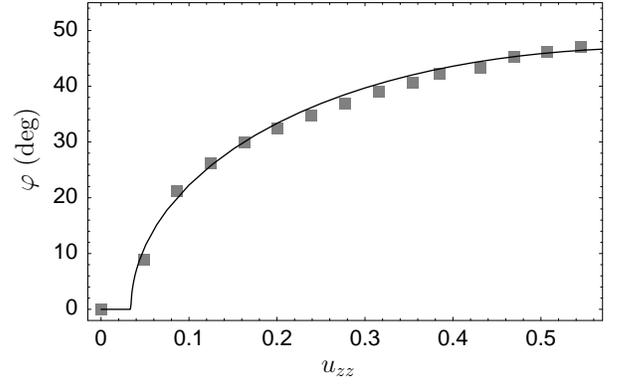}}
\caption{Dependence of the tilt angle $\varphi$ on
$u_{zz}$. The squares symbolize data taken from
Ref.~\cite{nishikawa_finkelmann_99}. The continuous curve
corresponds to our analytic result~(\ref{resVarphi}).}
\label{angleVsStrain}
\end{figure}

Finally, we calculate the tilt angle $\varphi_{\brm{n}}$ of the mesogens in
response to $u_{zz}$. The angle that the target-space director makes with the $z$ axis is, as depicted in Fig.~\ref{fig:strains}, 
\begin{equation}
\varphi_{\brm{n}} = \alpha - \gamma \, ,
\end{equation}
where $\alpha = \sin^{-1} {\tilde c}_x^0$. Like $\varphi$, $\varphi_{\brm{n}}$ vanishes as $\delta
u_{zz} \rightarrow 0$. In the vicinity of $\delta u_{zz} = 0$, its expansion
to order $(\delta u_{zz})^{3/2}$ is
\begin{equation}
\label{resVarphiDirector}
\varphi_{\brm{n}} = \left\{
\begin{array}{ccc}
0& \quad \mbox{for} \quad & u_{zz} < u_{zz}^c \, ,
\\
 A_{\brm{n}, 1} \, (\delta u_{zz})^{1/2} +  A_{\brm{n}, 3} \, (\delta u_{zz})^{3/2}
 & \quad \mbox{for} \quad& u_{zz} > u_{zz}^c ,
\end{array}
\right.
\end{equation}
with the value of $ A_{\brm{n}, 1}$ being determined entirely by the lowest order
expressions for ${\tilde c}_x^0$ [Eq.~(\ref{eq:c0x})] and $u_{xz}^0$
[Eq.~(\ref{eq:u0xz})],
\begin{equation}
 A_{\brm{n}, 1}  =  s + \frac{s \rho_c}{1 + u_{zz}^c} . 
\label{eq:B_1}
\end{equation}
$ A_{\brm{n}, 3}$ has contributions both from the expansions of $\sin^{-1}$ and
$\tan^{-1}$ to third order in their arguments, and from $(\delta
u_{zz})^{3/2}$ contributions to ${\tilde c}_x^0$ and $u_{xz}^0$,
which we have not calculated. From the BZGGZ and AD values we obtain, respectively, approximately $A_{\brm{n}, 1} \approx 85$ and $A_{\brm{n}, 1} \approx 90$ when $\varphi_{\brm{n}}$ is measured in degrees.

In their experiments, NF did not seek to produce curves of $\varphi_{\brm{n}}$ versus $u_{zz}$. From measuring the layer thickness as a function of $u_{zz}$ and by interpreting their X-ray reflection patterns, however, they conclude that the layer normal and the director remain parallel above the threshold strain and that the reduction of $Y_\parallel$ above the threshold stems form a strain induced breakdown of the smectic layers into pieces with local Sm$A$ order. Our theory, on the other hand, predicts that the angles $\varphi$ and $\varphi_{\brm{n}}$ are different but of the same order of magnitude, and that the reduction of $Y_\parallel$ results from shear and Sm$C$-like ordering produced by sample extension. If the difference between $\varphi$ and $\varphi_{\brm{n}}$ is small, it can be difficult to distinguish between Sm$A$ and Sm$C$ order experimentally. If this is the case in the experiments of NF, then the differences between the experimental findings and our theoretical predictions for the tilt angles may lie within the experimental error. 

Above, we have at several occasions briefly commented on the relation of our work to that of AW. A more detailed comparison of the two approaches can be found in the Appendix.

\section{Electroclinic effect in Sm$A^\ast$ elastomers}
\label{electroclinicEffect}
In a chiral smectic elastomer an external electric field $\brm{E}$
can couple to both the elastic and the orientational degrees of
freedom. First, we consider the former coupling. $\brm{E}$,
like the director  $\brm{n}$ and the layer
normal $\brm{N}$, is a vector in the target space and hence cannot be
coupled directly to the reference space tensor $\tens{u}$. By now,
of course, we know how to overcome this problem: we can use the
polar decomposition theorem to switch from $\brm{E}$ to its
reference space counterpart $\tilde{\brm{E}}$ via the
transformation $\tilde{\brm{E}}= \tens{O}^T \brm{E}$ and then
construct form $\tilde{\brm{E}}$ and $\tens{u}$ couplings that are
manifestly invariant under simultaneous rotations of the sample
and the field in the reference space. Since we are interested in
Sm$A^\ast$ elastomers, the couplings have to conform with uniaxial
symmetry in the reference space which leads us to
\begin{align}
\label{Eenergy1}
f_{E,1} = - v_1 \, \tilde{E}_a \hat{u}_{ab}  \tilde{E}_b - v_2 \,
u_{zz} \tilde{E}_z^2  \, ,
\end{align}
where $v_1$ and $v_2$ are coupling constants. In Eq.~(\ref{Eenergy1}) we have omitted a term of the type $u_{aa} \tilde{E}_b^2$ to keep our discussion as simple as possible. This omission does not affect our results qualitatively. 

Now we come to the coupling between the electric field and the orientational degrees
of freedom. In a chiral smectic a spontaneous polarization
$\brm{P}$ occurs if the director $\brm{n}$ and the layer normal
$\brm{N}$ are neither parallel nor perpendicular:
\begin{align}
\label{polarization}
\brm{P} \sim   \brm{n} \times \brm{N}  \, (\brm{n} \cdot \brm{N})
\, .
\end{align}
Thus, in the presence of $\brm{E}$, the elastic energy density of
the elastomer has an extra contribution
\begin{align}
\label{Eenergy2}
f_{E,2} \sim \brm{E} \cdot \brm{n} \times \brm{N} \,  (\brm{n}
\cdot \brm{N})    \, .
\end{align}
In order to combine $f_{E,2}$ with the remaining terms of the
elastic energy density we recast $f_{E,2}$ in the reference space
to obtain
\begin{align}
\label{Eenergy2ReferenceSpace}
f_{E,2} =  K\,  \tilde{\brm{E}} \cdot  \tilde{\brm{n}} \times
\tilde{\brm{N}} \,  ( \tilde{\brm{n}} \cdot  \tilde{\brm{N}}) \, ,
\end{align}
where $K$ is a coupling constant. The experimental findings by BZGGZ imply $K \sim 10^{-3}$ Pa(m/V), whereas the value of $K$ implied in the results by AD is one order of magnitude smaller, $K \sim 10^{-4}$ Pa(m/V).

In the typical experimental setup, the normal of the elastomeric
film is along the $z$ direction and the electric field $\brm{E}$ is along the
$y$ direction. If we assume that the experiments correspond to the
geometry shown in Fig.~\ref{fig:electrostriction}, then
$\Lambda_{zx}^0 =  0$ but $\Lambda_{xz}^0 \neq  0$, and the only
nonzero components of $\tens{\Lambda}$ are are $\Lambda_{xx}^0$,
$\Lambda_{yy}^0$, $\Lambda_{zz}^0$, and $\Lambda_{xz}^0$, implying
that that $O_{xy} = O_{yx} = O_{yz} = O_{zy} = 0$.  Thus ${\tilde
E}_i = O_{ij}^T E_j = E_i$, and $\tilde{\brm{E}} = (0, E, 0)$.  In
addition, ${\tilde N}_i = [g^{-1/2}]_{iz} [g^{-1}]_{zz}^{-1/2}
\approx (- u_{az}, 1 - \frac{1}{2} u_{az}^2 )$. With these expressions, 
the above coupling energy densities simplify to
\begin{subequations}
\begin{align}
f_{E,1} &= - v_1 \, \hat{u}_{yy} \, E^2   ,
\\
\label{Eenergy2simple}
f_{E,2} &= - K\,  E\, (\tilde{c}_x + u_{xz})\, .
\end{align}
\end{subequations}
In Eq.~(\ref{Eenergy2simple}) we have dropped higher order terms in
$\tilde{c}_a$ which are inconsequential for the leading behavior
since we are interested in the Sm$A^\ast$ and not in the Sm$C^\ast$
phase. For the same reason, we need, when we combine $f_{E,1}$ and
$f_{E,2}$ with the remaining parts of $f$, to keep in $f_c$ and
$f_{\text{coupl}}$ only terms up to second order in $\tilde{c}_a$.
To keep our discussion as simple as possible, we will focus in the
following on the incompressible limit. If strains are small, as we assume, we can implement this limit by setting $u_{ii} = 0$.  In accordance with the given geometry, we set $u_{xy} = u_{yz} =
{\tilde c}_y = 0$. Collecting all parts of the elastic energy
density we then obtain
\begin{align}
\label{completeEnWithField}
f &= f_{\text{uni}} + f_{\text{nonlin}} + f_{c} + f_{\text{coupl}} + f_{E,1} + f_{E,2}
\nonumber \\
&= \textstyle{\frac{1}{2}} \, (C_1+ C_4) \, u_{zz}^2 + 2 C_4
(u_{xx}^2 +  u_{xx} u_{zz})
 + C_5 \,u_{az}^2
\nonumber \\
&+\textstyle{\frac{1}{2}} \, r \,\tilde{c}_x^2 +
\textstyle{\frac{1}{2}} \, ( 2 \lambda_1 + \lambda_3) \,
\tilde{c}_x^2 \, u_{zz} - B_1 u_{zz} u_{xz}^2
\nonumber \\
&+ \lambda_3 \, \tilde{c}_x^2 \, u_{xx} +  \lambda_4 \, \tilde{c}_x
u_{xz} + \lambda_5 u_{zz} u_{xz} {\tilde c}_x
\nonumber \\
&- K\,  E \, (\tilde{c}_x + u_{xz})+ v_1 \, (u_{xx} + u_{zz}) \, E^2
,
\end{align}
where we have eliminated $u_{yy}$ by using $u_{ii} = 0$ and where we
dropped the $B_2 (u_{az}^2)^2$ term.

Next, we determine the equilibrium values of the strain components
and the c-director in the presence of the electric field.  The
equilibrium values ${\tilde c}_x^0$ and $u_{zx}^0$, which are at leading order linear
in $E$, are
\begin{subequations}
\label{equilibriumWithE}
\begin{align}
\tilde{c}_x^0 &= \frac{2 C_5 - \lambda_4}{2 C_5 r_R}  K E \, ,
 \\
u_{zx}^0 &=   - \frac{\lambda_4 -r }{ 2 \, C_5 \, r_R} \, K E =
\frac{r - \lambda_4}{2 C_5 - \lambda_4} {\tilde c}_x^0 \, ,
\end{align}
\end{subequations}
where $r_R = r - \lambda_4^2/(2 C_5)$. The diagonal components of $u_{ij}^0$ can also be calculated by minimizing the free energy of Eq.~(\ref{completeEnWithField}).  The
results are that $u_{zz}^0$ and $u_{xx}^0$ both have contributions
proportional directly to $E^2$ and contributions proportional to
$({\tilde c}_x^0)^2$. In addition $u_{zz}^0$ has a contribution
proportional to $- B (u_{xz}^0)^2$. All of these terms are proportional
to $E^2$.

The relative height change 
\begin{align}
\Delta h /  h_0 = (h  - h_0)/  h_0  =  u_{zz}^0
\end{align}
of the film is
\begin{align}
\label{hightChange}
\Delta h /  h_0 &=  -\frac{1}{C_1} [ \lambda_1 ({\tilde c}_x^0)^2 + \lambda_5
{\tilde c}_x^0 u_{xz}^0 - B_1 (u_{xz}^0 )^2 ] -\frac{v_1}{2 C_1} E^2
\nonumber \\
& \approx - \frac{1}{2} [ ({ \tilde c}_x^0 + u_{xz}^0 )^2 - 4
(u_{xz}^0 )^2 ] -\frac{v_1}{2 C_1} E^2  ,
\end{align}
where we used $B_{\text{sm}} \gg \mu, r_t$ to obtain the last form
(which sets $f_{\text{layer}}= 0$). The electrostriction coefficient is defined  via $\Delta h /  h_0 \equiv - a_{\text{el}} \, E^2$. Inserting Eq.~(\ref{equilibriumWithE}) into Eq.~(\ref{hightChange}), we obtain
\begin{align}
a_{\text{el}} & \approx  \frac{v_1}{2 C_1} + \frac{2(2p-1)}{(p+1)^2} \left( \frac{2
C_5- \lambda_4 }{2C_5 r_R}
K \right)^2 \nonumber \\
& = \frac{v_1}{2 C_1} +  \frac{1}{2}\frac{2 p-1}{p^2}
\frac{K^2}{r_t^2} .
\label{electrostrictionCoefficient}
\end{align}

Finally, we compare our theoretical findings to the experimental results. In their experiments, Lehmann {\em et al}.\ applied alternating
lateral voltage, $U_{\text{ac}} (t) = U_{\text{ac}} \cos (\omega
t)$, so that $E = E_{\text{ac}} (t) = U_{\text{ac}} (t)/d$, where
$d$ is the width of the sample, and they measured the first and
the second harmonic of $\Delta h$. Rewriting
Eq.~(\ref{hightChange}) as
\begin{align}
\label{hightChangeHarmonic}
\Delta h =  \frac{h_0 \,  a_{\text{el}} \,  U_{\text{ac}}^2}{4\, d^2} \,  [1 +
\cos (2\omega t) ]+ \cdots \, ,
\end{align}
our model predicts that the amplitude $\Delta h_2$ of the second
harmonic is
\begin{align}
\label{hightChangeHarmonicAmp}
\Delta h_2 =  \frac{h_0 \,  a_{\text{el}} \,  U_{\text{ac}}^2}{4\, d^2}  \, .
\end{align}
In Fig.~\ref{fig:electrostrictionComparison} our
result~(\ref{hightChangeHarmonicAmp}) is compared to the
experimental data of Lehmann {\em et al}. Taking the values $h_0 =
75 \pm 5 \, \mbox{nm}$ and $d = 1\, \mbox{mm}$ of
Ref.~\cite{lehmann&Co_01} and setting $a_{\text{el}} =7.147\times 10^{-14} \,
(\mbox{m}/ \mbox{V})^{2}$, we obtain an excellent fit between our theory and the experimental curve for the second harmonic. Likewise, we can fit our theory easily to the experimental curves of K\"{o}ler {\em et al}.~\cite{KohlerZen2005}. We refrain here to show the corresponding curves in order to save space. Beyond producing fitting curves, we can extract from our theory an estimate for the value of $a_{\text{el}}$. Let us start with the first term in the second line of Eq.~(\ref{electrostrictionCoefficient}). The coefficient $v_1$ should be of order $\epsilon_0 \chi_0$ where $\epsilon_0 = 8.85
\times 10^{-12}$ Pa \, (m/V)$^2$ is the permeability of the vacuum and
$\chi_0 \approx 2.5$ is the electric susceptibility.  With these
values, the term containing $v_1$ is of order $2 \times 10^{-11}/4
\times 10^7\approx 0.5 \times 10^{-18}$ (m/V)$^2$, which is orders of magnitude smaller than the experimental values.  Now, we turn to the second term in the second line of Eq.~(\ref{electrostrictionCoefficient}).  In accord with this term, the experiments of Lehmann {\em et al}.\ and K\"{o}ler {\em et al}.\ found the largest values of $a_{\text{el}}$ for temperatures near $T_c$. Based on the BZGGZ and the AD values, we estimate $r_t$ to be of the order $r_t \sim 10^4$ Pa and $r_t \sim 10^5$ Pa, respectively, for $T$ within a range of a few degrees about $T_c$. Recalling the experimental values for $K$ and that $p\approx 1$, were are led to  $a_{\text{el}} \approx 5 \times 10^{-15} (\mbox{m}/ \mbox{V})^{2}$ for the BZGGZ values and $a_{\text{el}} \approx  \times 10^{-18} (\mbox{m}/ \mbox{V})^{2}$ for the AD values. Perhaps not surprisingly, the latter estimate, being based on experimental data for liquid smectics,  turns out poor. The estimate $a_{\text{el}} \approx 5 \times 10^{-15} (\mbox{m}/ \mbox{V})^{2}$ is in much better agreement with the electrostriction experiments. It is about one order of magnitude smaller than the value $a_{\text{el}} =7.147\times 10^{-14} \, (\mbox{m}/ \mbox{V})^{2}$ obtained by fitting the data of Lehmann {\em et al}.\ and it is of the same order of magnitude  as the value $a_{\text{el}}  =(1 \pm 0.2) \times 10^{-15}$ (m/V)$^2$ quoted in Ref.~\cite{KohlerZen2005}.
\begin{figure}
\centerline{\includegraphics[width=8.4cm]{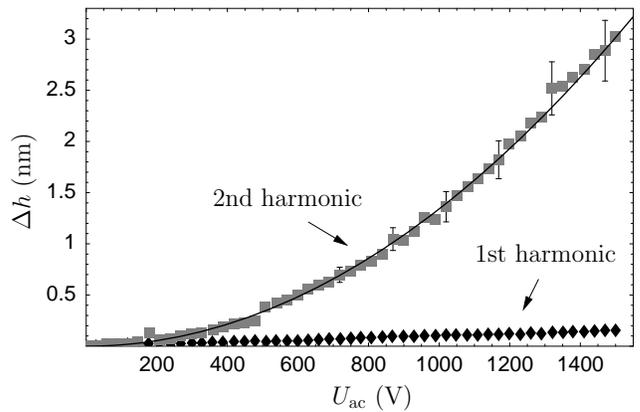}}
\caption{Hight change of a Sm$A^\ast$ elastomer in response to
a alternating lateral voltage $U_{\text{ac}} (t) = U_{\text{ac}}
\cos (\omega t)$ as a function of the amplitude $U_{\text{ac}}$.
The data points are taken from Ref.~\cite{lehmann&Co_01}, where a
sample of thickness $h_0 = 75 \pm 5 \, \mbox{nm}$ and width $d =
1\, \mbox{mm}$ was used and the frequency of the voltage was
$\omega = 133 \, \mbox{Hz}$. There is a small piezoelectric
contribution, i.e., a contribution that is linear in
$U_{\text{ac}}$ or respectively the amplitude of the electric
field $E_{\text{ac}} (t) = U_{\text{ac}} (t)/d$. In contrast, the
contribution proportional to the square of the field, i.e., the
effect of electrostriction, at the second harmonic is large. The
continuous curve stems from our theoretical
result~(\ref{hightChangeHarmonicAmp}) with $a_{\text{el}}=7.147\times 10^{-14}
\,(\mbox{m}/ \mbox{V})^{2}$.}
\label{fig:electrostrictionComparison}
\end{figure}

\section{Concluding remarks}
\label{concludingRemarks}
We have investigated the behavior of monodomain Sm$A$ elastomers
under strains $u_{zz}$ along the smectic layer normal. When
exceeding a threshold value $u_{zz}^c$, these strains induce a
transition to a Sm$C$-like state. This effect is accompanied by a
tilt of the mesogenic component that sets in at the same
threshold. Due to the transition, the Young's modulus
$Y_\parallel$ is qualitatively different from that of conventional
uniaxial elastomers: the slope of the stress-strain curve changes from high constant value for
$u_{zz} < u_{zz}^c$ to a much lower constant value for $u_{zz} <
u_{zz}^c$.  Our results for the stress-stain behavior as well as for the layer tilt
agree nicely with the available experimental data.

Moreover, we studied the electroclinic effect in Sm$A^\ast$
elastomers in an external, lateral electric field. Here, it is the
external field that induces a transition to a Sm$C^\ast$-like
state. Since this state is sheared in the plane perpendicular to
the field, the height of a sample is lower than in the initial
Sm$A^\ast$ phase. Our theory finds, in absolute agreement with the
experimental evidence, that his height change is proportional to
the square of magnitude of the external field, i.e., that the
mechanism at work is the so-called electrostriction. Our numerical estimate for the electrostriction coefficient is in accord with the available experimental values. 

\begin{acknowledgments}
This work was supported in part by  National Science Foundation
under grant DMR 0404670 (TCL).
\end{acknowledgments}

\appendix
\section*{Note on the work of Adams and Warner}
\label{app:adams_warner}
Here we will discuss the connection between the AW theory and our
theory for Sm$A$'s in some more detail to understand the agreement
in the results more systematically. Moreover, we will generalize the AW theory to allow for relative till between the director and the layer normal.

\subsection{Comparison to the work of Adams and Warner}
\label{app:comparison_adams_warner}
AW extended the neoclassic
model of rubber elasticity by formulating an elastic energy
density
\begin{align}
\label{AWenergy}
f_{\text{AW}} = \textstyle{\frac{1}{2}} \mu \mbox{Tr} \left[
\tens{\Lambda} \, \tens{\ell}_0 \,  \tens{\Lambda}^T \tens{\ell}^{-1}
\right] + \textstyle{\frac{1}{2}} B (d/d_0 -1)^2 \, .
\end{align}
The first term on the right hand side of Eq.~(\ref{AWenergy}) is
the usual trace formula of the neoclassic model. $\mu$ is the shear
modulus. 
\begin{align}
\tens{\ell}_0 = \tens{\delta} + (p-1) \nv_0 \nv_0 
\end{align}
is the so-called shape tensor describing the
conformations of polymeric chains before deformation and
\begin{align}
\tens{\ell}^{-1}  =  \tens{\delta} - (1-p^{-1}) \nv \nv 
\end{align}
is its counterpart after a deformation has been
applied. As in the main text, $p$ denotes the anisotropy ratio of the uniaxial state, which we assume to be prolate, $p>1$. The AW model assumes incompressibility, i.e., the
deformation tensor is subject to the constraint $\det
\tens{\Lambda} =1$. The new element in the theory of AW, the
second term on the right hand side of Eq.~(\ref{AWenergy}),
describes changes in the spacing of smectic layers. $B$ is a layer
compression modulus. $d_0$ and $d$ are respectively the layer
spacing before and after deformation which are related via
\begin{align}
\frac{d}{d_0} = \frac{1}{\big|(\tens{\Lambda}^{-1})^T \brm{n}_0 \big|} \, .
\end{align}
In the following, we chose $\nv_0 = (0,0,1)$, and we set $\nv = (\sin\theta, 0 ,\cos\theta)$. Assuming a deformation tensor of the form shown in Eq.~(\ref{formOfLambda}), the elastic energy density becomes
\begin{align}
& f  =  \frac{1}{2} \mu \bigg\{ \Lambda_{zz}^2 + \Lambda_{xx}^2 +
\Lambda_{xx}^{-2} \Lambda_{zz}^{-2} + p^{-1} \Lambda_{zx}^2 \nonumber\\
& (1-p^{-1})(p \Lambda_{zz}^2   -
\Lambda_{xx}^2 + \Lambda_{zx}^2) \sin^2 \theta \nonumber\\
&  - 2 ( 1 - p^{-1})\Lambda_{xx} \Lambda_{zx} \sin \theta \cos \theta \} \nonumber\\
& + b \bigg( \frac{\Lambda_{zz} \Lambda_{xx}}{\sqrt{\Lambda_{xx}^2 +
\Lambda_{zx}^2}} - 1 \bigg)^2 \bigg\} ,
\label{eq:full_f}
\end{align}
where $\Lambda_{yy}$ has  been eliminated via the incompressibility constraint, and where $b = B/\mu$.
For studying the response of an Sm$A$ elastomer to an imposed
deformation $\Lambda_{zz}$, AW assume that $\brm{n}$ and $\brm{N}$ are locked, which amounts to setting
\begin{equation}
\sin\theta = -\frac{\Lambda_{zx}}{\sqrt{\Lambda_{xx}^2 +
\Lambda_{zx}^2}} , \qquad \cos\theta =
\frac{\Lambda_{xx}}{\sqrt{\Lambda_{xx}^2 + \Lambda_{zx}^2}} .
\end{equation}
Then, the AW elastic energy density becomes
\begin{align}
\label{AWenergyzz}
f_{\text{AW}} &=\frac{1}{2} \mu \bigg\{ \Lambda_{xx}^2 +
\frac{1}{\Lambda_{xx}^2 \Lambda_{zz}^2} + \Lambda_{zx}^2  +
\frac{(\Lambda_{xx}^2 + p \Lambda_{zx}^2)
\Lambda_{zz}^2}{\Lambda_{xx}^2 +  \Lambda_{zx}^2}
\nonumber \\
&+ b \bigg(  \frac{\Lambda_{xx} \Lambda_{zz}}{\sqrt{\Lambda_{xx}^2
+  \Lambda_{zx}^2}}  - 1 \bigg)^2 \bigg\}  .
\end{align}

In order to make contact to our Lagrangian theories, we re-express
the components of the deformation tensor in terms of the
components of the strain tensor. From the definition of the strain
tensor and (\ref{formOfLambda}) it is straightforward to see that
the respective components are related by
\begin{subequations}
\label{strainDefRel}
\begin{align}
\Lambda_{yy} &= \sqrt{1 + 2 u_{yy}}\, ,
\\
\Lambda_{zz} &= \sqrt{1 + 2 u_{zz}}\, ,
\\
\Lambda_{zx} &= u_{xz}/\Lambda_{zz}  \, ,
\\
\Lambda_{xx} &= \sqrt{1 + 2 u_{xx} - \Lambda_{zx}^2} \, .
\end{align}
\end{subequations}
Inserting Eq.~(\ref{strainDefRel}) into the elastic energy
density~(\ref{AWenergyzz}) to eliminate the $\Lambda_{ij}$ and
then expanding in powers of the $u_{ij}$ we obtain
\begin{align}
\label{AWenergyzzInStrains}
f_{\text{AW}} &=\mu \bigg\{ \frac{3}{2}  + \frac{4+b}{2} u_{zz}^2
+ 2 u_{zz} u_{xx} + 2  u_{xx}^2 + \frac{p}{2} u_{zx}^2
\nonumber \\
&-  \frac{4+b}{2}  u_{zz} u_{zx}^2 - (1+p)  u_{xx} u_{zx}^2 +
\frac{4+b}{2} u_{zx}^4 \bigg\} \, ,
\end{align}
where we have dropped higher order terms which are inconsequential
for the argument here.

Next, we revisit elastic energy density $f$ in
Eq.~(\ref{completeEn}). To obtain a model in terms of
strain only, we integrate $\tilde{c_a}$ out of $f$. This procedure
leads to
\begin{align}
f &= f_{\text{uni}} + D_1\, u_{zz} u_{az}^2 + D_2 \,  u_{ii}
u_{az}^2 + D_3 \, \hat{u}_{ab} u_{az} u_{bz}
\nonumber \\
&+ G  \, (u_{az}^2)^2 ,
\end{align}
with $C_5$ renormalized to $C_{5}^R$ [see Eq.~(\ref{C5ren})] in $ f_{\text{uni}}$,
with $D_m = \lambda_m \lambda_4^2/r$ for $m=1,2,3$, and $G = (g/4
+ r/2)\lambda_4^4/r^4$. Then, we carry out the following steps:
(i) we implement the incompressible limit via the constraint
$u_{ii}=0$ so that the terms featuring $C_2$, $C_3$, and $D_2$ are
absent, (ii) we eliminate $u_{yy}$ with help of $u_{ii}=0$, and
(iii) we assume a deformation tensor of the
form~(\ref{formOfLambda}) so that $u_{xy}=u_{yz}=0$. Then,  $f$
reduces to
\begin{align}
\label{f2uniReduced}
f &=\textstyle{\frac{1}{2}} (C_1 + C_4) u_{zz}^2+  2 C_4 u_{zz}
u_{xx} +  2 C_4 u_{xx}^2 + C_{5}^{R} u_{zx}^2
\nonumber \\
&+(D_1 + \textstyle{\frac{1}{2}} D_3)  u_{zz} u_{zx}^2 + D_3
u_{xx} u_{zx}^2 + G u_{zx}^4  \, ,
\end{align}
which has, besides lacking the inconsequential constant term,
exactly the same form as $f_{\text{AW}}$. Comparison of the
individual terms in Eqs.~(\ref{AWenergyzzInStrains}) and
(\ref{f2uniReduced}) reveals that the elastic constants in the 2
models are related by $C_1 = \mu (3+b)$, $C_4 = \mu$, $C_{5}^R = p
\mu$, $D_1 = \mu (p-3-b)/2$, $D_3 = - \mu (1+p)$, and $G= \mu
(4+b)/8$. Note that AW work with the values $b=5$ and $p=2$ which
corresponds to negative values of $D_1$ and $D_3$.

The effective elastic energy density in terms of $u_{zz}$ only
produced by Eq.~(\ref{f2uniReduced}) is essentially the same as
the incompressible limit of Eq.~(\ref{effectiveCompleteModel}).
Thus, in the end, the AW model and our models lead, at least to
leading order, to the same stress-strain curve, cf.\ Fig.~\ref{SmAstressStrain}.

\subsection{Extending the model of Adams and Warner}
\label{app:extending_adams_warner}
Now we generalize the AW model by avoiding the assumption that the layer normal and the director are locked. The full elastic energy density~(\ref{eq:full_f}) rather than the AW elastic energy density~(\ref{AWenergyzz}) will be the vantage point for the following considerations. In the remainder, we abbreviate $\Lambda \equiv \Lambda_{zz}$ for notational simplicity.

We can reduce Eq.~(\ref{eq:full_f}) to a function of $\Lambda$,
$\Lambda_{xx}$, and $\Lambda_{zx}$ only by minimizing over $\theta$.
The result is
\begin{subequations}
\begin{align}
\sin\theta& =  \sqrt{\frac{1}{2}}\left[1 - \sqrt{1+
\frac{\mathcal{B}^2}{\mathcal{A}^2}} \right] \\
\cos \theta & =  \sqrt{\frac{1}{2}}\left[1 + \sqrt{1+
\frac{\mathcal{B}^2}{\mathcal{A}^2}} \right] ,
\end{align}
\end{subequations}
where
\begin{subequations}
\begin{align}
\mathcal{A} & =  p \Lambda^2 + \Lambda_{zx}^2 - \Lambda_{xx}^2 \\
\mathcal{B}& =  2 \Lambda_{xx} \Lambda_{zx} .
\end{align}
\end{subequations}
The free energy minimized over $\theta$ is then
\begin{align}
f_m & = \frac{1}{2} \mu \bigg\{ (1+p) \Lambda^2 + (1+p^{-1})
\Lambda_{xx}^2 +2 \Lambda^{-1} \Lambda_{xx}^{-2} \nonumber \\
&  + (1+p^{-1})\Lambda_{zx}^2 \nonumber \\
&  - (1-p^{-1}) \sqrt{(r \Lambda^2 - \Lambda_{xx}^2 +
\Lambda_{zx}^2)^2 + 4 \Lambda_{xx}^2 \Lambda_{zx}^2} \Big] \nonumber
\\
& + b \bigg( \frac{\Lambda \Lambda_{xx}}{\sqrt{\Lambda_{xx}^2 +
\Lambda_{zx}^2}} - 1 \bigg)^2  \bigg\}.
\end{align}
To determine instabilities toward the development of $\Lambda_{zx}$,
we expand $f_m$ to second order in $\Lambda_{zx}$:
\begin{align}
f_m & \approx  \frac{1}{2} \mu \big[ p \Lambda^2 + \Lambda_{xx}^2 +
\Lambda^{-2} \Lambda_{xx}^{-2}] + b ( \Lambda -1 )^2  \big]
\nonumber \\
&  + \frac{1}{2} \, \mu \left(\frac{\Lambda^2 - \Lambda_{xx}^2}{p
\Lambda^2 - \Lambda_{xx}^2} - 2 b
\frac{\Lambda(\Lambda-1)}{\Lambda_{xx}^2} \right) \Lambda_{zx}^2
\end{align}
Now we minimize this function over $\Lambda_{xx}$ at $\Lambda_{zx} =
0$.  The result is
$\Lambda_{xx} = \Lambda^{-1/2}$.
Thus, when $\Lambda = 1$, $\Lambda_{xx} = 1$, $\Lambda_{xx}^2 =
\Lambda^2$, and the coefficient of $\Lambda_{zx}$ is zero, i.e.,
there is no restoring force for infinitesimal shear strains in this
model of smectics.  The origin of this behavior is the invariance
of both terms in the free energy with respect to rotations.
Semi-soft terms need to be added to endow the smectic layers with a
preferred direction relative to the crosslinked matrix.  Next, we 
consider what happens when $\Lambda>1$.  The coefficient of
$\Lambda_{zx}^2$ becomes
\begin{equation}
A_{zx} = - \frac{1}{2} \, \mu (\Lambda -1 ) \left( b \Lambda^2 - 
\frac{\Lambda^2+ \Lambda + 1}{p \Lambda^3 - 1} \right) .
\end{equation}
This terms is negative for all $\Lambda>1$ because $b \gg 1$.
Thus, in the absence of a semi-soft term, there is no threshold for
the production of shear strain in response to an imposed strain
along the $z$ direction.

As we have just seen, there is an is an instability to transfer
shear for all $\Lambda >1$ in the original AW theory if the nematic
director is allowed to relax. Thus, the model elastic energy density~(\ref{eq:full_f}) is insufficient to produce stress-strain curves as measured by NF.
The problem is that this model is
invariant with respect to simultaneous rotations of the smectic
layers, the nematic director, and $\Lambda$ in the target space.  To
break this invariance, we introduce the semisoft energy
\begin{align}
f_{\rm semi}&= \frac{1}{2} \mu \alpha {\rm Tr}[(\tens{\delta} -
\nv_0 \nv_0) \tens{\Lambda}^T \nv \nv \tens{\Lambda}]  \\
& =  \frac{1}{2} \mu \alpha (\Lambda_{xx} \sin\theta + \Lambda_{zx}
\cos \theta )^2 .
\end{align}
The complete energy is then
\begin{equation}
f_T = f + f_{\rm semi} .
\end{equation}
Expanding to second order in $\theta$, we obtain
\begin{align}
& f  =  \frac{1}{2} \mu \bigg\{ \Lambda^2 + \Lambda_{xx}^2 +
\Lambda_{xx}^{-2} \Lambda^{-2} + (p^{-1}+ \alpha) \Lambda_{zx}^2
 \nonumber\\
&  - 2 (1-p^{-1} - \alpha)\Lambda_{xx} \Lambda_{zx}  \theta 
 \nonumber\\
& + [(1-p^{-1})(p \Lambda^2  -
\Lambda_{xx}^2 + \Lambda_{zx}^2)+ \alpha (\Lambda_{xx}^2 - \Lambda_{zx}^2)] \theta^2 
\nonumber\\
& + b \bigg( \frac{\Lambda \Lambda_{xx}}{\sqrt{\Lambda_{xx}^2 +
\Lambda_{zx}^2}} - 1 \bigg)^2 \bigg\}.
\label{eq:full_fT}
\end{align}
Minimizing over $\theta$, we obtain
\begin{equation}
\theta = \frac{(1-p^{-1} - \alpha)\Lambda_{xx}
\Lambda_{zx}}{(1-p^{-1})(p \Lambda^2  - \Lambda_{xx}^2 +
\Lambda_{zx}^2)+ \alpha (\Lambda_{xx}^2 - \Lambda_{zx}^2)}
\end{equation}
and a relaxed free energy to harmonic order in $\Lambda_{zx}$ of
\begin{align}
f_m  & \approx  \frac{1}{2} \mu[ p \Lambda^2 + \Lambda_{xx}^2 +
\Lambda^{-2} \Lambda_{xx}^{-2}]+ b ( \Lambda -1 )^2 ]
\nonumber \\
&  + \frac{1}{2} \, \mu \bigg[ \frac{(p-1)(\Lambda^2 - \Lambda_{xx}^2)+ p
\alpha ((p-1) \Lambda^2 + \Lambda_{xx}^2)} {(p-1)(p \Lambda^2 -
\Lambda_{xx}^2) + \alpha \Lambda_{xx}^2} \nonumber \\
& - 2 b \frac{\Lambda(\Lambda-1)}{\Lambda_{xx}^2} \bigg]
\Lambda_{zx}^2 .
\end{align}
The term of order $0$ in $\Lambda_{zx}$ has the same form as it did
in the absence of the semi-soft term.  Thus, as before,
$\Lambda_{xx} = \Lambda^{-1/2}$, and the coefficient of
$\Lambda_{zx}^2$ is
\begin{align}
A &= \frac{\mu}{2} \bigg[\frac{(p-1)(\Lambda^3 - 1) + \alpha p [(p-1)
\Lambda^3 + 1]}{(p-1)(p \Lambda^3 - 1) + \alpha p}
\nonumber \\
& - 2 b \Lambda^2 ( \Lambda-1) \bigg] .
\end{align}
If $\Lambda = 1 + u_{zz}$, then to lowest order in $u_{zz}$,
\begin{align}
A &= \frac{\mu}{2}\bigg[ \frac{\alpha p^2}{(p-1)^2 + \alpha p} - 2 bu 
\nonumber \\
&+ \frac{3 (p-1)  (p (\alpha
   -1)+1)^2  }{\left((p-1)^2+p \alpha \right)^2} u 
\bigg] ,
\end{align}
and thus there is a nonzero the critical value of the strain $u_{zz}$,
\begin{equation}
u_{zz}^c =  \frac{p^2 \alpha \left[(1-p)^2+p \alpha \right] }{2 b \left[(1-p)^2+p \alpha \right] -3 (p-1) [p (\alpha -1)+1]^2} ,
\end{equation}
in accord with the experiments of NF and our Lagrangian theory. 
Note that this critical strain is linearly proportional to $\alpha$ at small $\alpha$ and vanishes
as $\alpha \rightarrow 0$ in agreement with our previous observation
that the elastomer is unstable to shear at infinitesimal strain along $z$ if it is soft rather than semi-soft.
Having $u_{zz}^c$, we can proceed to calculate the entire stress-strain curve. The result is in full qualitative agreement with the experimental curves and the predictions of our Lagrangian theory. To save space, we leave the remaining steps as an exercise to the reader.

\end{document}